# Zero-forward Kerker scattering via synthesized complex-frequency excitation

Jianfei Liu[1,†], Zheng Gong[1,2,†], Ao Li[1], Hongsheng Chen[1], Stefan Rotter[2,*], and Xiao Lin[1,*]

[†]*These authors contributed equally to this work.*

[*]*Corresponding authors: xiaolinzju@zju.edu.cn (X. Lin); stefan.rotter@tuwien.ac.at (S. Rotter).*

[1]*State Key Laboratory of Extreme Photonics and Instrumentation, Zhejiang Key Laboratory of Intelligent Electromagnetic Control and Advanced Electronic Integration, College of Information Science & Electronic Engineering, Zhejiang University, Hangzhou 310027, China.*

[2]*Institute for Theoretical Physics, Vienna University of Technology (TU Wien), A–1040 Vienna, Austria.*

**All objects illuminated with light inevitably cast a shadow – a universal phenomenon encapsulated in the fundamental property that all passive systems with plane-wave illumination feature a non-zero forward scattering amplitude. Since Kerker's landmark 1983 paper, considerable effort has been devoted to overcoming this limitation and achieving the elimination of forward scattering – an objective now widely known as the zero-forward Kerker scattering. However, this objective is fundamentally restricted to active systems, requiring either physical gain in materials or virtual gain in the excitation source. Despite advances in active materials science and non-Hermitian photonics, the experimental realization of zero-forward Kerker scattering still remains an open challenge. Here, we show, both theoretically and experimentally, how zero-forward Kerker scattering can be effectively synthesized by a weighted superposition of readily accessible real-frequency responses. Our synthetic recipe builds on creating an artificial pole at a complex frequency that prevails over inherent scattering poles. These findings not only unlock a realistic experimental framework for non-Hermitian light-matter interactions, but also hold technological relevance for non-invasive sensing and imaging.**

## Introduction

Light scattering is ubiquitous in nature[1–4]. One of its basic manifestations is that physical objects inevitably cast shadows when illuminated. The core signature underlying this everyday observation is the *forward scattering* amplitude. As governed by the renowned optical theorem[5–7], the forward scattering amplitude accurately quantifies how much a scatterer extinguishes a single incident plane wave through both absorption and scattering. Consequently, for any passive system that inherently absorbs or scatters light, the forward scattering amplitude is intrinsically non-zero. Indeed, it has become an established principle in photonics that forward scattering from passive systems excited by a single plane wave cannot be eliminated[5,8,9].

In fact, the quest to manipulate or suppress forward scattering has garnered extensive interest[10–17], owing to its pivotal role in encapsulating the total optical response and encoding key material properties. The foundational milestone in this pursuit is Kerker's seminal 1983 paper[10], in which hypothetical magnetic spheres were theoretically proposed to eliminate backward scattering or partially suppress forward scattering. The cornerstone concept of optically induced artificial magnetism in metamaterials[11,12,18] eventually led to the successful experimental realization of the *first* Kerker scattering with zero backward scattering[13,14,19]. While zero-backward Kerker scattering is now well established and has found broad relevance in electromagnetic and acoustic systems, with applications such as directional scattering control[20,21], achieving the *second* Kerker scattering with zero forward scattering poses a far more demanding challenge. Specifically, active systems with optical gain have theoretically been proven to be mandatory for zero-forward Kerker scattering[9,17,22,23].

To date, the experimental realization of zero-forward Kerker scattering remains unresolved, constrained by the difficulty in the implementation of active systems and the excitation and/or measurement of light scattering therein. On the one hand, physical gain media are unavoidably accompanied by intrinsic fabrication imperfections, large-area inconsistency, and temporal instability[24,25]. On the other hand, recent advances in complex-frequency excitations reveal that non-time-harmonic types of excitation may provide an alternative route to implement an active system via so-called virtual gain[26–31]. Yet, complex-frequency waves and fields exhibiting temporally decaying profiles are generally difficult to obtain and/or measure, particularly in non-one-dimensional scattering setups.

Here, we circumvent the challenges of realizing zero-forward Kerker scattering in active systems by leveraging complex-frequency synthesis. Complex-frequency synthesis is a technique that reconstructs the optical response to the excitation at a target complex frequency via a weighted integral over the easily accessible real-frequency spectrum in experiments[32]. Fundamentally, this technique builds upon the integration of Cauchy's integral formula and physical causality[33,34] – the same mathematical framework that historically led to the discovery of optical sum rules[35], fundamental limits[36], and the Kramers-Kronig relations[37]. On the other hand, complex-frequency synthesis has rapidly emerged as a transformative tool across diverse fields, including super-resolution imaging[32,38,39], ultrasensitive biosensing[40], physics-informed machine learning[41], and electron microscopy[42]. However, it remains an open question whether this complex-frequency synthesis technique is valid in intricate scattering systems, and more importantly, whether it can enable the experimental realization of zero-forward Kerker scattering.

Here we show that this goal can indeed be achieved. Specifically, we create an artificial pole at a target complex frequency characterized by a vanishing forward scattering amplitude and a much more pronounced contribution than that of inherent scattering poles. On this basis, we experimentally realize zero forward scattering by measuring and superposing the scattering responses across the real-frequency spectrum. Moreover, we find a causality-governed spatiotemporal evolution of the synthesized scattered wavepacket, with a transition from the transient to steady states. Our findings emphasize the "unreasonable effectiveness of mathematics"[43] in retrieving beyond-material-limit responses even in resonant scattering systems. Moreover, our synthetic recipe potentially paves the way towards wavefront shaping[2] in the temporal domain, virtually incorporating gain and loss into scattering systems, holding technological relevance across a wide variety of applications, ranging from scattering spectroscopy and microscopy[44], bio-imaging[45], to spectrophotometry[46], chemical analysis[47] and quantum thermometry[48].

## Results

Figure 1 conceptually illustrates zero-forward Kerker scattering via complex-frequency excitation. Our schematic setup follows standard scattering theory[4,5] (see also supplementary Figs. S1 and S2). A plane wave with a working frequency $\omega$ travels along the $z$ direction, and is incident upon the all-dielectric spherical scatterer centered at the coordinate origin in vacuum (Fig. 1a). The incident electric field takes the form of $\overline{E}_{\text{inc}} = \hat{x}E_0e^{-i\omega t}$ in Cartesian coordinates. Without loss of generality, we study the spherical scattering properties within the $yz$ plane, for which $\phi = 90°$ in $(r, \theta, \phi)$ spherical coordinates, where $\theta$ is the polar angle between the wavevectors of incident and scattered light, and $\phi$ is the azimuthal angle. Confined to this plane, the scattered field features only a $\hat{\phi}$-component (i.e., $\hat{\phi}E_{\text{s}}(\omega)e^{-i\omega t}$), where $E_{\text{s}}(\omega)$ is evaluated as the difference between the total and incident fields, namely

$$E_{\text{s}}(\omega) = E_{\text{tot}}(\omega) - E_{\text{inc}}(\omega) \tag{1}$$

Correspondingly, the differential scattering cross section of interest, $\sigma_{\text{sca}}^{\text{d}}(\theta)$, is related to the total scattering cross section $\sigma$ via

$$\sigma_{\text{sca}}^{\text{d}}(\theta) = \left.\frac{\text{d}^2\sigma}{\text{d}\phi\text{d}\theta}\right|_{\phi=90°} \tag{2}$$

Generally, light scattering under time-harmonic excitations in passive systems is always characterized by *non-zero* forward scattering (Fig. 1a). Specifically, for any real-valued frequency, the forward scattering cross section is strictly positive, namely $\sigma_{\text{sca}}^{\text{d}}(0°) > 0, \forall\omega = \omega_0 \in \mathbb{R}$. Yet as shown in Fig. 1b,c, by incorporating non-Hermiticity and allowing for a complex-valued frequency of light, *zero* forward scattering can emerge in the complex-frequency plane, i.e., $\sigma_{\text{sca}}^{\text{d}}(0°) = 0$ at $\omega = \omega_{\text{c}} = \omega_0 - i \cdot |\text{Im}(\omega_{\text{c}})| \in \mathbb{C}$. The specific value of $\omega_{\text{c}}$ is determined by mapping $\sigma_{\text{sca}}^{\text{d}}(\omega, \theta = 0°)$ in the complex-frequency plane and searching for its zeros (see also supplementary Fig. S3), that belongs to a discrete set $\Omega_{\text{c}}$, namely

$$\omega_{\text{c}} \in \Omega_{\text{c}}, \quad \text{where } \Omega_{\text{c}} \coloneqq \{\omega \in \mathbb{C} | \sigma_{\text{sca}}^{\text{d}}(\omega, \theta = 0°) = 0\}. \tag{3}$$

In other words, to satisfy equation (3) and thereby realize zero forward scattering, the passive scatterer must be explicitly excited at a complex zero $\omega = \omega_{\text{c}}$. The existence of such zeros in the lower complex-frequency plane has been theoretically established in recent literature[22,23]. In practice, this means that the incident wave must be designed with a specific temporally decaying profile corresponding to $\omega_c = \omega_0 - i \cdot |\mathrm{Im}(\omega_c)|$. However, the experimental observation has remained a long-standing challenge, due to the practical inaccessibility of scattering excitation and/or measurements involving temporal decay. In fact, directly accessing complex-frequency responses in actual experiments is currently limited to simple geometries[27,31] and remains unresolved for standard scattering setups.

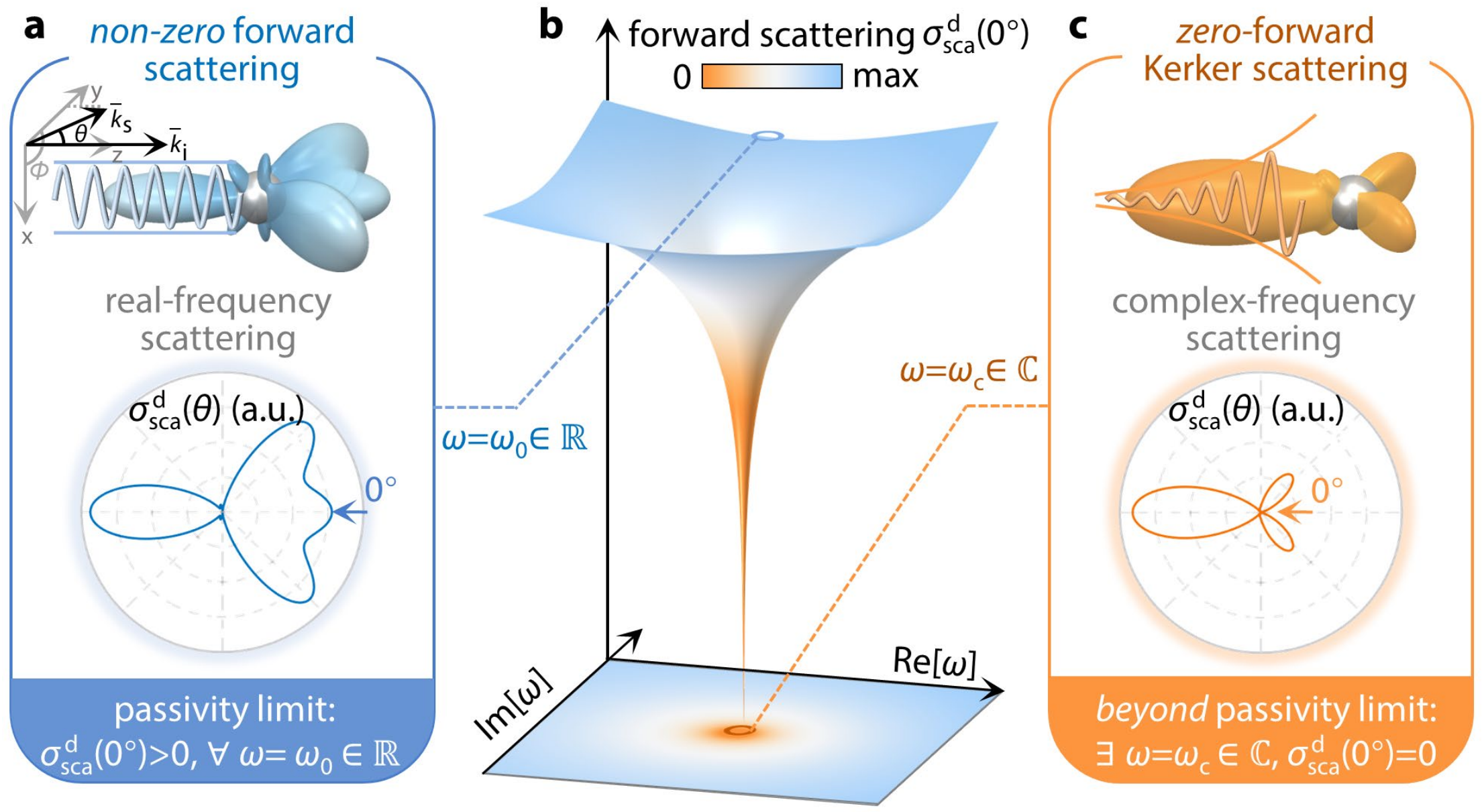

**Fig. 1 | Conceptual illustration of zero-forward Kerker scattering via complex-frequency excitation. a**, Non-zero forward scattering for any real frequency. The incident light interacts with a passive scatterer at a working frequency $\omega$ when traveling along the $z$ direction. Conventionally, for any time-harmonic, monochromatic plane-wave incidence, meaning that the working frequency is real-valued (i.e., $\omega = \omega_0 \in \mathbb{R}$), the forward scattering cannot be eliminated, $\sigma_{\mathrm{sca}}^{\mathrm{d}}(0°) > 0$. **b**, Forward scattering cross section $\sigma_{\mathrm{sca}}^{\mathrm{d}}(0°)$ mapped in the complex-frequency plane. **c**, Zero-forward Kerker scattering at a complex frequency. For non-time-harmonic, complex-frequency plane-wave incidence, meaning that the working frequency is complex-valued (e.g., $\omega = \omega_c = \omega_0 - i \cdot |\mathrm{Im}(\omega_c)| \in \mathbb{C}$), the forward scattering can vanish, $\sigma_{\mathrm{sca}}^{\mathrm{d}}(0°) = 0$. Unless otherwise specified, throughout the main text, $\sigma_{\mathrm{sca}}^{\mathrm{d}}(\theta)$ is normalized to $\max\left[\sigma_{\mathrm{sca}}^{\mathrm{d}}(\theta), \theta\right]$ of each scattering pattern, respectively; the spherical scatterer has a relative permittivity of $\varepsilon_r = 10.2$ and a diameter of $d = 31.75$ mm. The scatterer size is comparable to the wavelength of incident light, namely $d \approx \lambda_0 = 2\pi c/\mathrm{Re}(\omega_c)$, where $c$ is the light speed in vacuum and $\omega_c/2\pi = (9.98 - 0.13i)$ GHz.

To address the challenge of realizing complex-frequency excitation in scattering systems, we propose a universal synthetic framework in Fig. 2, relying solely on the commonly accessible real-frequency measurements. This framework includes three fundamental types of scattered fields.

First, the *ideal* scattered field $E_{\mathrm{s}}^{\mathrm{ideal}}$ represents the scattering response to the excitation at a complex frequency $\omega = \omega_{\mathrm{c}}$ where zero forward scattering can be achieved (Fig. 2a). By definition, $E_{\mathrm{s}}^{\mathrm{ideal}}$ can be expressed as the residue of an analytic complex function $E_{\mathrm{s}}(\omega)e^{-i\omega t}/(\omega - \omega_{\mathrm{c}})$, namely

$$E_{\mathrm{s}}^{\mathrm{ideal}} = E_{\mathrm{s}}(\omega_{\mathrm{c}})e^{-i\omega_{\mathrm{c}}t} = \mathrm{Res}\left[\frac{E_{\mathrm{s}}(\omega)e^{-i\omega t}}{\omega - \omega_{\mathrm{c}}}, \omega_{\mathrm{c}}\right] \tag{4}$$

where the residue of $f$ for its simple pole $z_0$ is defined as $\mathrm{Res}[f, z_0] = \lim_{z \to z_0}(z - z_0)f(z), z \in \mathbb{C}$[33]. Notably, the pole at the target complex frequency $\omega_{\mathrm{c}}$ is created artificially, in the sense that it is mathematically introduced by the synthesis kernel via a non-physical term $1/(\omega - \omega_{\mathrm{c}})$, rather than an inherent scattering pole.

Second, the *synthesized* scattered field $E_{\mathrm{s}}^{\mathrm{syn}}$ accounts for a weighted integral over the experimentally accessible real-frequency spectrum (Fig. 2b), namely

$$E_{\mathrm{s}}^{\mathrm{syn}} = \frac{1}{2\pi i}\int_{+\infty}^{-\infty} \mathrm{d}\omega \frac{E_{\mathrm{s}}(\omega)e^{-i\omega t}}{\omega - \omega_{\mathrm{c}}} \tag{5}$$

Finally, the *residual* scattered field $E_{\mathrm{s}}^{\mathrm{res}}$ is defined as the sum of residues at the inherent scattering poles $\omega_{\mathrm{pole}}^{i}$ $(i = 1, 2, 3, \ldots)$ (Fig. 2c), namely

$$E_{\mathrm{s}}^{\mathrm{res}} = \sum_{\mathrm{all}\ i} \mathrm{Res}\left[\frac{E_{\mathrm{s}}(\omega)e^{-i\omega t}}{\omega - \omega_{\mathrm{c}}}, \omega_{\mathrm{pole}}^{i}\right] \tag{6}$$

The scattering poles at the complex frequencies $\omega_{\mathrm{pole}}$, defined by the root of the equation $1/E_{\mathrm{s}}(\omega_{\mathrm{pole}}^{i}) = 0$, are not artificially created but physically originate from the scattering system itself.

The intrinsic relation among the ideal, synthesized, and residual scattered fields can be readily analyzed in the complex-frequency plane (Fig. 2d). Based on causality, an integration contour $\boldsymbol{C}$ enclosing both the artificial pole and scattering poles is drawn on the real axis and in the lower half of the complex-frequency plane. Based on this enclosed contour and Cauchy's integral formula, one has $\frac{1}{2\pi i}\oint_C \mathrm{d}\omega \frac{E_\mathrm{s}(\omega)e^{-i\omega t}}{\omega-\omega_\mathrm{c}} = E_\mathrm{s}^\mathrm{syn} + \frac{1}{2\pi i}\int_{C_R} \mathrm{d}\omega \frac{E_\mathrm{s}(\omega)e^{-i\omega t}}{\omega-\omega_\mathrm{c}} = E_\mathrm{s}^\mathrm{ideal} + E_\mathrm{s}^\mathrm{res}$ , where $C_R \coloneqq \{\omega|\omega = R\cdot e^{i\varphi}, \varphi \in [0,\pi], R \to \infty\}$ is a semicircle of infinite radius on the lower complex-frequency plane. Since the integral over the semicircular arc $\frac{1}{2\pi i}\int_{C_R} \mathrm{d}\omega \frac{E_\mathrm{s}(\omega)e^{-i\omega t}}{\omega-\omega_\mathrm{c}}$ vanishes by Jordan's lemma[33], we arrive at the key principle for synthesizing complex-frequency scattering, namely

$$E_\mathrm{s}^\mathrm{ideal} = E_\mathrm{s}^\mathrm{syn} - E_\mathrm{s}^\mathrm{res} \tag{7}$$

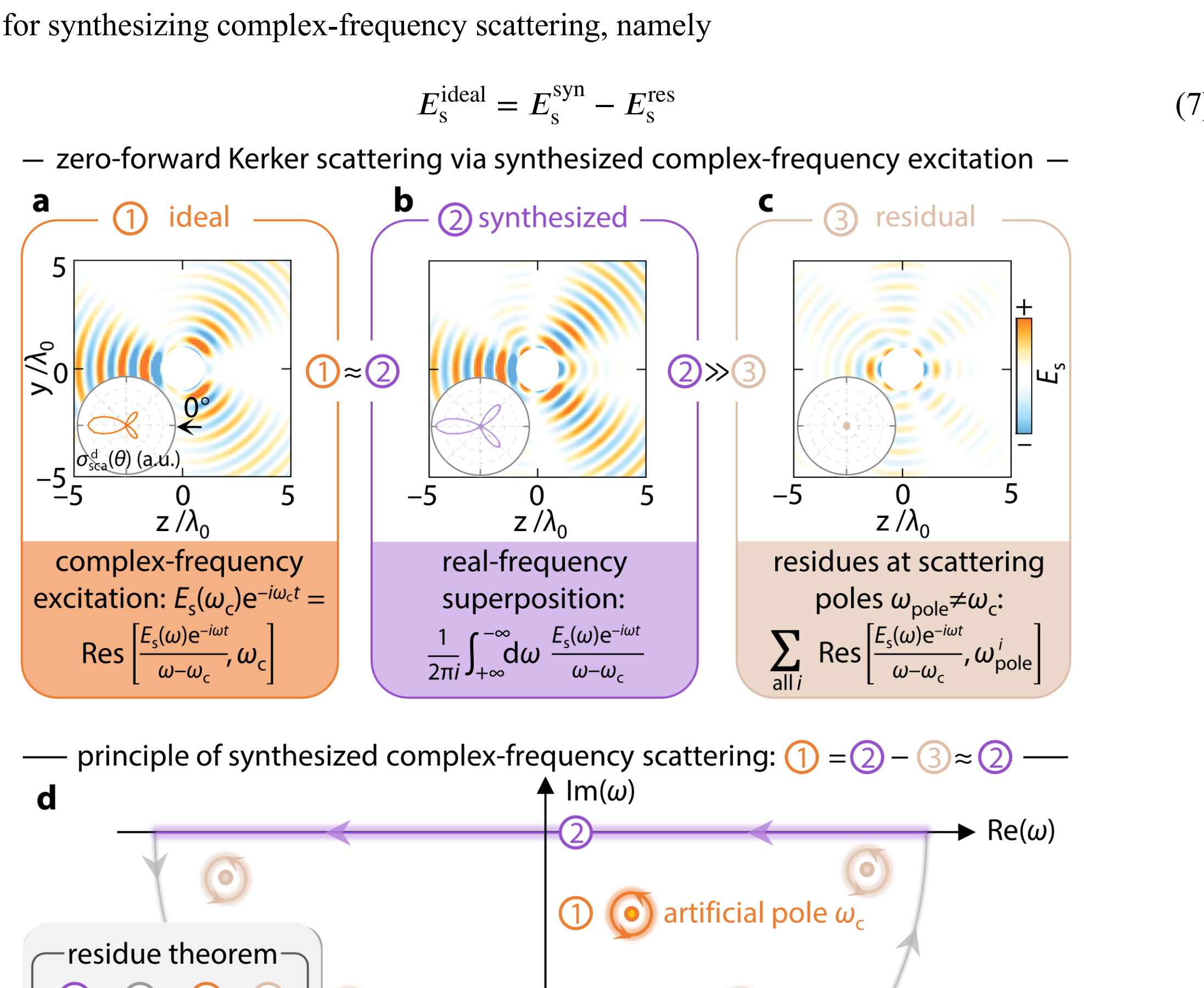

**Fig. 2 | Synthesizing zero-forward Kerker scattering via the superposition of real-frequency responses.** The principle of synthesizing zero-forward Kerker scattering at a complex frequency hinges on the relationship among three fundamental types of scattered fields in **a-c**, namely the ideal (①) and synthesized (②) complex-frequency scattered fields, and the residual (③) scattered field. The residual term ③ represents the discrepancy between the ideal and synthesized complex-frequency scattering,

namely ①=②−③ as shown in **d**. For the insets in **a-c**, $\sigma_{\mathrm{sca}}^{\mathrm{d}}(\theta)$ is normalized to a common factor (i.e., the maximum value $\max\left[\sigma_{\mathrm{sca}}^{\mathrm{d}}(\theta),\theta\right]$ in **b**).

Upon close inspection of Fig. 2a-c , the validity of synthesized complex-frequency scattering can be readily guaranteed, if the contribution from the inherent scattering poles is negligible compared with the artificial pole (Fig. 2a-c), namely

$$E_{\mathrm{s}}^{\mathrm{ideal}} \cong E_{\mathrm{s}}^{\mathrm{syn}}, \quad \text{if } |E_{\mathrm{s}}^{\mathrm{res}}| \ll \left|E_{\mathrm{s}}^{\mathrm{ideal}}\right| \tag{8}$$

For this purpose, we carefully choose the artificial pole at a target complex frequency $\omega_{\mathrm{c}}$ from the set $\Omega_{\mathrm{c}}$. Notably, the artificial pole introduces a resonance in the synthesized real-frequency field spectrum, centered at $\mathrm{Re}(\omega_{\mathrm{c}})$ and characterized by a bandwidth $2|\mathrm{Im}(\omega_{\mathrm{c}})|$ in the real-frequency synthesized field spectrum. When this artificial resonance is chosen to avoid spectral overlap with those pronounced scattering resonances, the contributions from the scattering poles are suppressed by the factor $1/(\omega-\omega_{\mathrm{c}})$ (see supplementary Fig. S4). As a result, the synthesized complex-frequency scattering faithfully recovers the features of ideal complex-frequency scattering, for example, the eliminated forward scattering (Fig. 2a,b).

At this point, the synthesis of complex-frequency excitation is straightforward for the experimental realization of zero-forward Kerker scattering (Fig. 3). The experimental setup comprises two primary workflows: acquiring data from real-frequency measurements, and numerically synthesizing complex-frequency responses in post-processing (Fig. 3a). A probe is controlled to scan the region of interest before and after the scatterer is placed, to acquire the incident and total field distributions, respectively (i.e., $E_{\mathrm{inc}}(\omega)$ and $E_{\mathrm{tot}}(\omega)$ ). Based on these experimental measurements, the complex-frequency scattering is numerically synthesized using equation (5), namely

$$E_{\mathrm{s}}^{\mathrm{syn}} \approx \sum_{\omega\in\Omega_{\mathrm{r}}} \frac{E_{\mathrm{s}}(\omega)e^{-i\omega t}}{2\pi i(\omega_{\mathrm{c}}-\omega)}\Delta\omega \tag{9}$$

where $E_{\mathrm{s}}(\omega)$ is given by equation (1), $\Omega_{\mathrm{r}}$ is a set of discrete real frequencies at which the measurement is conducted and $\Delta\omega$ is the frequency step. Throughout the main text, we use a synthesis bandwidth of 1.6 GHz and synthesis step of 0.05 GHz.

In principle, as long as each real-frequency scattering measurement is accurate, the experimentally synthesized complex-frequency scattering can be successfully obtained. While all the real-frequency scattering responses are characterized by a non-zero forward scattering at $\theta = 0°$ (Fig. 3b), the zero-forward Kerker scattering emerges at a synthesized complex frequency (Fig. 3c). Our synthetic strategy is particularly versatile as it is not limited to three-dimensional spherical scattering geometries, but also facilitates the realization of cylindrical zero-forward Kerker scattering (see supplementary Fig. S5).

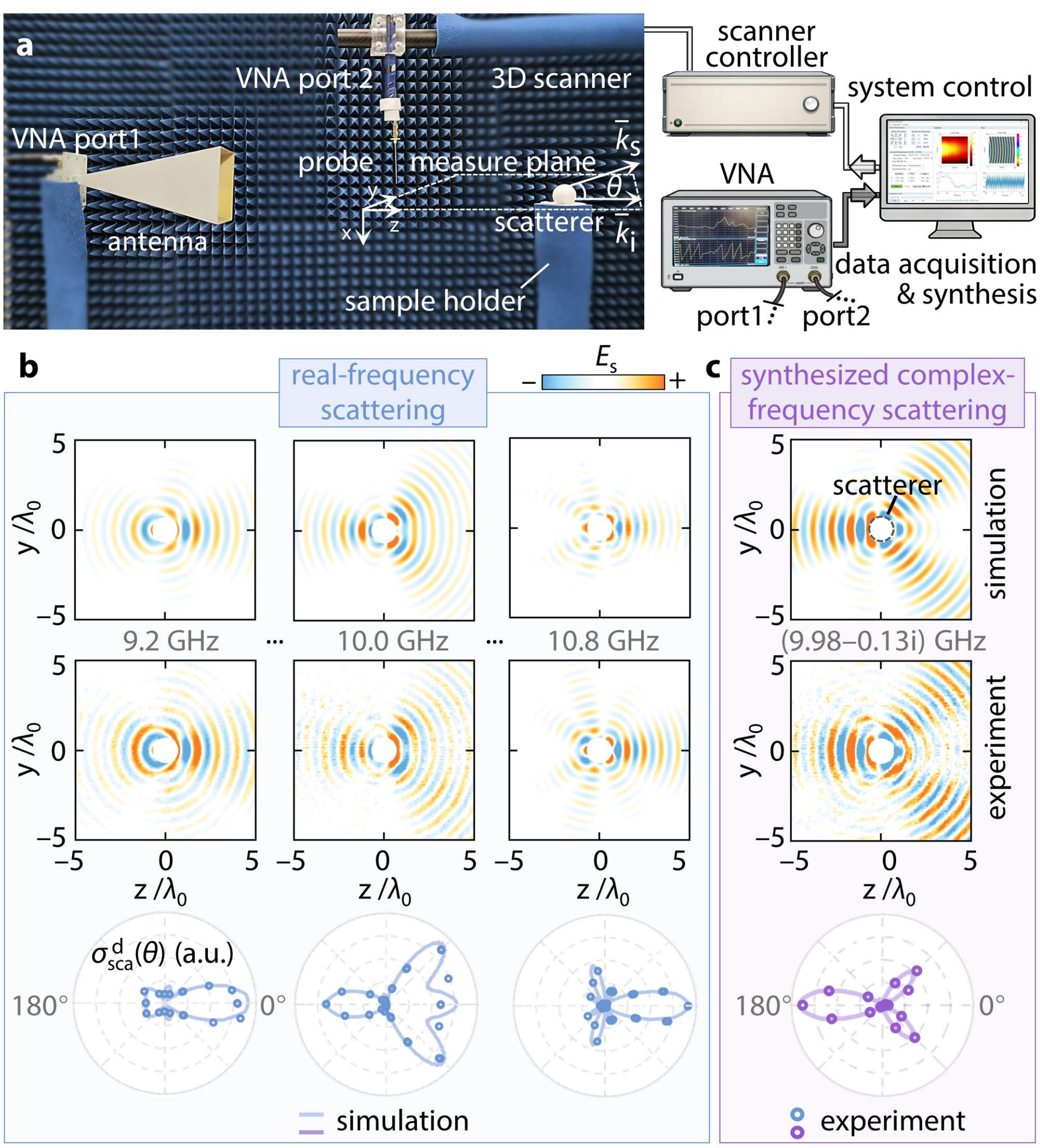

**Fig. 3 | Experimental realization of zero-forward Kerker scattering via synthesized complex-frequency excitation. a**, Experimental setup. **b**,**c**, Simulated and measured fields and differential scattering cross sections. The scattered field was measured by scanning a probe across the observation plane (i.e., the *yz* plane) across the real-frequency spectrum (e.g., 9.2~10.8 GHz). On this basis, through the numerically weighted superposition of real-frequency measurements, the desired complex-frequency scattering response is effectively synthesized.

To provide deeper insight, we examine the spatiotemporal evolution of the synthesized complex-frequency scattering (Fig. 4).

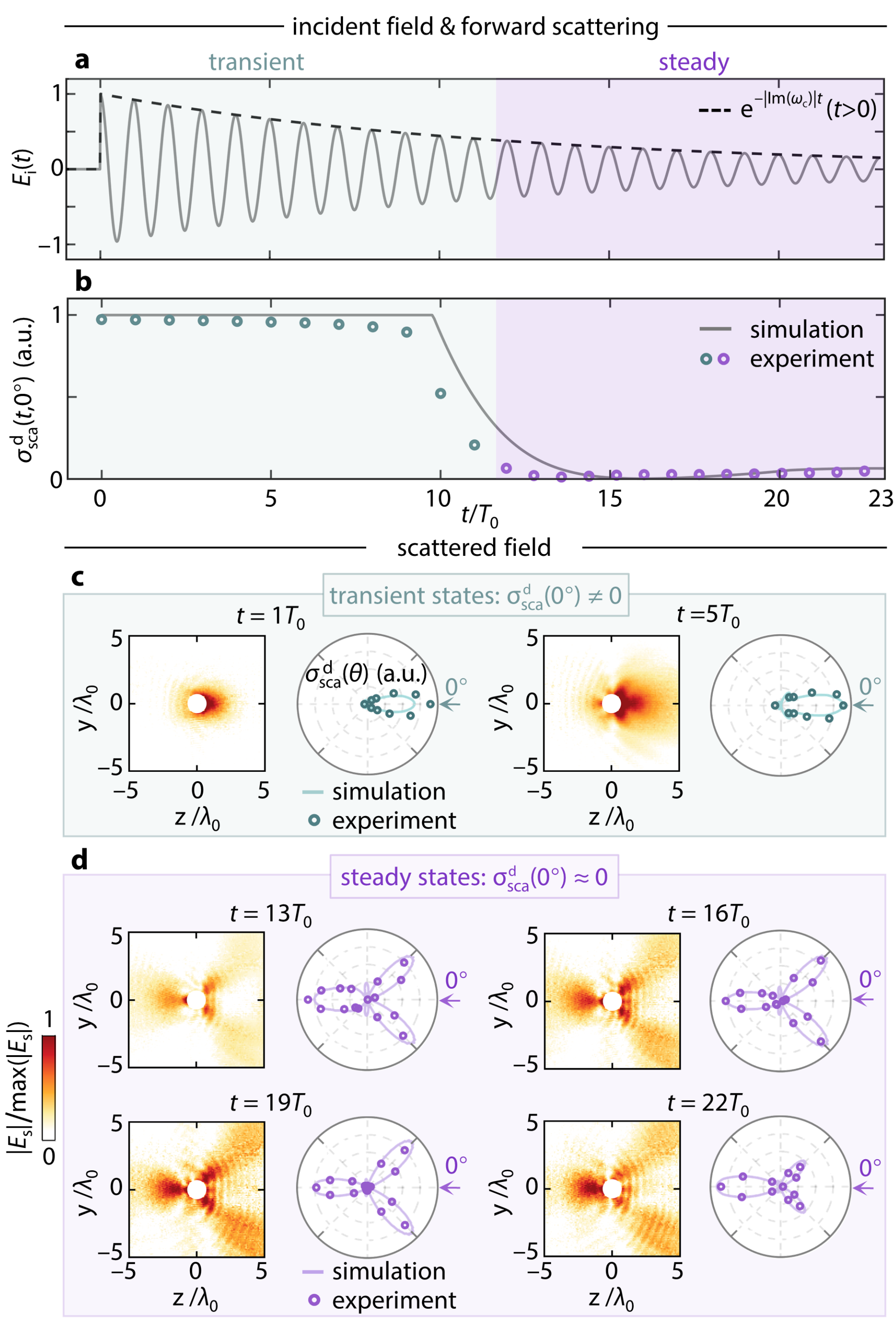

**Fig. 4 | Spatiotemporal evolution of synthesized complex-frequency scattering. a**,**b**, Evolution of incident-field profile at the scatterer centroid and forward scattering cross section over time. **c**,**d**, Evolution of the synthesized scattered wavepacket from transient to steady states. In **b**, $\sigma^{\mathrm{d}}_{\mathrm{sca}}(t, 0°)/\max\left[\sigma^{\mathrm{d}}_{\mathrm{sca}}(t,\theta),\theta\right]$ is used for normalization; meanwhile, a small synthesis error $\Delta(t) \lesssim 20\%$ is

used to distinguish the transient and steady states, where $\Delta(t) = \max\left[\sigma_{\text{sca}}^{\text{d,res}}(t,\theta),\theta\right] / \max\left[\sigma_{\text{sca}}^{\text{d,ideal}}(t,\theta),\theta\right]$ is defined as the ratio between the residual contribution from the inherent scattering poles $\max\left[\sigma_{\text{sca}}^{\text{d,res}}(t,\theta),\theta\right]$ and the ideal complex-frequency scattering response $\max\left[\sigma_{\text{sca}}^{\text{d,ideal}}(t,\theta),\theta\right]$.

The elimination of forward scattering is restricted to a time interval, outside which the forward scattering cross section would still deviate from zero. The main reason is the causality principle, which sets a fundamental time stamp at which the time decaying complex-frequency wave could start, to guarantee energy converge. Specifically, a temporal cutoff is mandatory to obey causality; for example, on the scatterer centroid, there is no incident signal activated before $t = 0$ (Fig. 4a). Consequently, an initial transient period (e.g., $t/T_0 \in [0,12]$) emerges during which the incident complex-frequency wavepacket has not fully interacted with the scatterer, where $T_0$ is the temporal period of light in vacuum. During this transient period, the forward scattering cross section is non-zero, namely $\sigma_{\text{sca}}^{\text{d}}(0°) \neq 0$ (Fig. 4b,c). As time increases, the synthesized complex-frequency scattering evolves into steady states (e.g., $t/T_0 \in [12,23]$) (Fig. 4b,d), characterized by a zero forward scattering cross section, namely $\sigma_{\text{sca}}^{\text{d}}(0°) \approx 0$. This regime marks our main observation of zero-forward Kerker scattering with steady-state zero-forward-scattering property. In short, the causality principle governs the spatiotemporal evolution of the synthesized complex-frequency scattering, driving it from transient to steady states (see supplementary movie S1).

Before closing, we emphasize that the synthesized zero-forward Kerker scattering faithfully reproduces the response under physical complex-frequency excitation (see Fig. S6) and originates from destructive interference among higher-order multipoles[49] (see Fig. S7). The underlying reason is that the residual contribution from the inherent scattering poles remains much smaller than that from the synthesized artificial pole, provided that the synthesis frequency, bandwidth, and frequency step are properly chosen (see Fig. S8). More importantly, for a properly chosen synthesis frequency, zero-forward Kerker scattering exhibits a certain degree of robustness against variations in experimental imperfections, scatterer properties, and material loss (see Figs. S9-S12).

## Conclusion

In conclusion, we have revealed a universal synthetic framework for light scattering operating at previously inaccessible complex working frequencies, based on which the zero-forward Kerker scattering is experimentally realized. More broadly, our synthetic framework serves as a practical platform for non-Hermitian photonics, facilitating the observation of intricate gain-loss interplay that is otherwise challenging and/or costly to realize[50]. Looking ahead, our results position complex-frequency synthesis as a powerful paradigm for reshaping wave scattering by introducing virtual gain and loss. Future efforts could focus on applying these unique features to complex *light* fields, such as spacetime wavepackets[51] and topologically structured light[52]. Moreover, extending the synthesized complex-frequency scattering from single-scatterer to multiple-scatterer regime (e.g., in random *media*) holds significant promise for imaging and computing in disordered systems[53]. Finally, our framework offers far-reaching implications for particle-matter interactions; for example, it potentially bridges the gap between theory and experiment in *free-electron* or *dipolar* interactions with gain media[54], paving the way for the energy gain spectroscopy[42].

## Author contribution

J.F., Z.G., and X.L. conceived the idea; J.F. and Z.G. performed the calculation and experiment; A.L., H.C., S.R., and X.L. helped to analyze the data; J.F., Z.G., S.R., and X.L. wrote the paper with input from all the other authors; S.R. and X.L. supervised the project.

## Conflict of interest

The authors have no conflicts to disclose.

## Data availability

All theoretical and numerical findings can be reproduced based on the information in the article and/or supplementary sections. Experimental raw data would be available on http://xxxx.

## Acknowledgement

X.L. acknowledges the support partly from National Key Research and Development Program of China under Grant No. 2025YFF0514901, the National Natural Science Foundation of China (NSFC) under Grants Nos. W2543013, 62475227, and W2641023, and Zhejiang Provincial Natural Science Foundation of China under Grant No. LR26F050002. H.C. acknowledges the support from the National Natural Science Foundation of China (NSFC) under Grant Nos. U25A20520 and 62475228, and the Key Research and Development Program of the Ministry of Science and Technology under Grants Nos. 2022YFA1404704 and 2022YFA1405201. S.R. acknowledges support by the Austrian Science Fund (FWF) through project META-INCOME [10.55776/PIN7240924]. Z.G. acknowledges support from Zhejiang University through the Qiushi Flying Eagle Program.

## Methods

**Experimental setup.** The standard horn antenna WR90 (BJ-100) with a gain of ≥ 20 dBi serves as the source operating in the frequency range 8.0~12.5 GHz; the open-ended semi-rigid coaxial cable with an exposed inner conductor and a diameter of 0.5 mm is used as the probe; a 3672B Vector Network Analyzer (Ceyear) is used for data acquisition; A three-dimensional scanning scattering microscopy system (LINBOU Nearfield Technology) with a spatial resolution of 0.1 mm is used; high-purity samples (≥ 99%) composed of Alumina ($Al_2O_3$) are used.

**Derivation of complex-frequency scattering theory.** We provide the derivation of the complex-frequency scattering theory in supplementary section S1. The complex-frequency scattering theory is in good agreement with COMSOL Multiphysics simulations, as demonstrated by the field distributions and radiation patterns in Figs. S1 and S2. On this basis, we map the forward scattering response in the complex-frequency plane in Fig. S3.

**More discussion on the artificial pole for zero-forward Kerker scattering.** We provide more details on creating an artificial pole at a target complex frequency in supplementary section S2 and Fig. S4.

**Synthesized zero-forward Kerker scattering by a cylinder.** We apply the synthesized complex-frequency framework to two-dimensional cylindrical scatterers in supplementary section S3. Based on real-frequency measurements, the zero-forward Kerker scattering by a cylinder is effectively realized at a synthesized complex frequency in Fig. S5.

**More discussion on synthesized zero-forward Kerker scattering.** We provide more discussion on synthesized zero-forward Kerker scattering in supplementary section S4, including the relationship among scattering responses under real-frequency and different types of complex-frequency excitation in Fig. S6, the multipolar decomposition of synthesized zero-forward Kerker scattering in Fig. S7, the sensitivity of zero-forward Kerker scattering to synthesis frequency, synthesis bandwidth, and frequency step in Fig. S8, experimental error estimates in Fig. S9, and the influence of synthesis frequency, scatterer properties, and material loss on synthesized zero-forward Kerker scattering in Figs. S10-S12.

Supplementary Information for

# Zero-forward Kerker scattering via synthesized complex-frequency excitation

Jianfei Liu, Zheng Gong, Ao Li, Hongsheng Chen, Stefan Rotter, and Xiao Lin

## The PDF file includes:

Supplementary Text and Figs. S1 to S12

## Other Supplementary Materials for this manuscript include the following:

Movie S1 for spatiotemporal evolution of synthesized zero-forward Kerker scattering

## Guide To the Supplementary Sections

# S1 Derivation of complex-frequency Mie scattering theory

In this section, we obtain a general formulation of scattering quantities under complex-frequency plane-wave excitations. Then, we map the forward scattering response in the complex-frequency plane.

## S1.1 Complex-frequency Mie scattering theory

In this subsection, we provide the analytical solution to complex-frequency scattered fields. On the basis of field distribution, we use Fourier transform to extract far-field radiation patterns.

### S1.1.1 Field distribution

Consider a complex-frequency plane wave travelling in $z$ direction with an electric field of $\overline{E}_{\text{inc}} = \hat{x} \cdot E_0 e^{ikz-i\omega t}$, where $\omega \in \mathbb{C}$ is the working frequency and $k = k_z = \omega/c$. A spherical isotropic dielectric particle of radius $a = d/2$ is located at the origin of the spherical coordinates $(r, \theta, \phi)$ where $\theta$ is the polar angle between the wavevectors of incident and scattered light, and $\phi$ is the azimuthal angle. Correspondingly, the scattered field $\overline{E}_{\text{s}} = \left(\hat{r} \cdot E_r + \hat{\theta} \cdot E_\theta + \hat{\phi} \cdot E_\phi\right) e^{-i\omega t}$ is given by[S1,S2]:

$$E_r = \frac{-iE_0 \cos\phi}{k^2 r^2} \sum\nolimits_{n=1}^{\infty} (2n+1) a_n(\omega) \widehat{H}_n^{(1)}(kr) P_n^1(\cos\theta) \tag{S1}$$

$$E_\theta = \frac{E_0 \cos\phi}{kr} \left( \sum\nolimits_{n=1}^{\infty} \frac{2n+1}{n(n+1)} \left[ -ia_n(\omega) \widehat{H}_n^{(1)\prime}(kr) \tau_n(\cos\theta) + b_n(\omega) \widehat{H}_n^{(1)}(kr) \pi_n(\cos\theta) \right] \right) \tag{S2}$$

$$E_\phi = \frac{E_0 \sin\phi}{kr} \left( \sum\nolimits_{n=1}^{\infty} \frac{2n+1}{n(n+1)} \left[ ia_n(\omega) \widehat{H}_n^{(1)\prime}(kr) \pi_n(\cos\theta) - b_n(\omega) \widehat{H}_n^{(1)}(kr) \tau_n(\cos\theta) \right] \right) \tag{S3}$$

where $\pi_n(\cos\theta) = P_n^1(cos\,\theta)/\sin\theta$ and $\tau_n(\cos\theta) = \mathrm{d}P_n^1(cos\,\theta)/\mathrm{d}\theta$, with $P_n^1$ being the associated Legendre polynomial of the first order, and $\widehat{H}_n^{(1)}$ denotes the Riccati-Bessel function of the third kind[S1,S2]; by enforcing the boundary condition the complex-frequency Mie coefficients $a_n(\omega)$ and $b_n(\omega)$ are determined:

$$a_n(\omega) = (-i)^{-n} \frac{-\sqrt{\varepsilon_\text{r}} \hat{J}_n'(ka) \hat{J}_n(k_s a) + \sqrt{\mu_\text{r}} \hat{J}_n(ka) \hat{J}_n'(k_s a)}{\sqrt{\varepsilon_\text{r}} \widehat{H}_n^{(1)\prime}(ka) \hat{J}_n(k_s a) - \sqrt{\mu_\text{r}} \widehat{H}_n^{(1)}(ka) \hat{J}_n'(k_s a)} \tag{S4}$$

$$b_n(\omega) = (-i)^{-n} \frac{-\sqrt{\varepsilon_\text{r}} \hat{J}_n(ka) \hat{J}_n'(k_s a) + \sqrt{\mu_\text{r}} \hat{J}_n'(ka) \hat{J}_n(k_s a)}{\sqrt{\varepsilon_\text{r}} \widehat{H}_n^{(1)}(ka) \hat{J}_n'(k_s a) - \sqrt{\mu_\text{r}} \widehat{H}_n^{(1)\prime}(ka) \hat{J}_n(k_s a)} \tag{S5}$$

where $k_\text{s} = \omega\sqrt{\varepsilon_\text{r}\mu_\text{r}}/c$ is the wave number defined with the scatterer, $\varepsilon_\text{r}$ and $\mu_\text{r}$ are the relative permittivity and permeability of the sphere, and $\hat{J}_n$ denotes the Riccati-Bessel functions of the first kind.

As seen from equations (S1)-(S5), the complex-frequency scattering theory simply follows conventional scattering theory, by extending the working frequency from real domain to complex domain. Its validity is verified by a COMSOL simulation (Fig. S1).

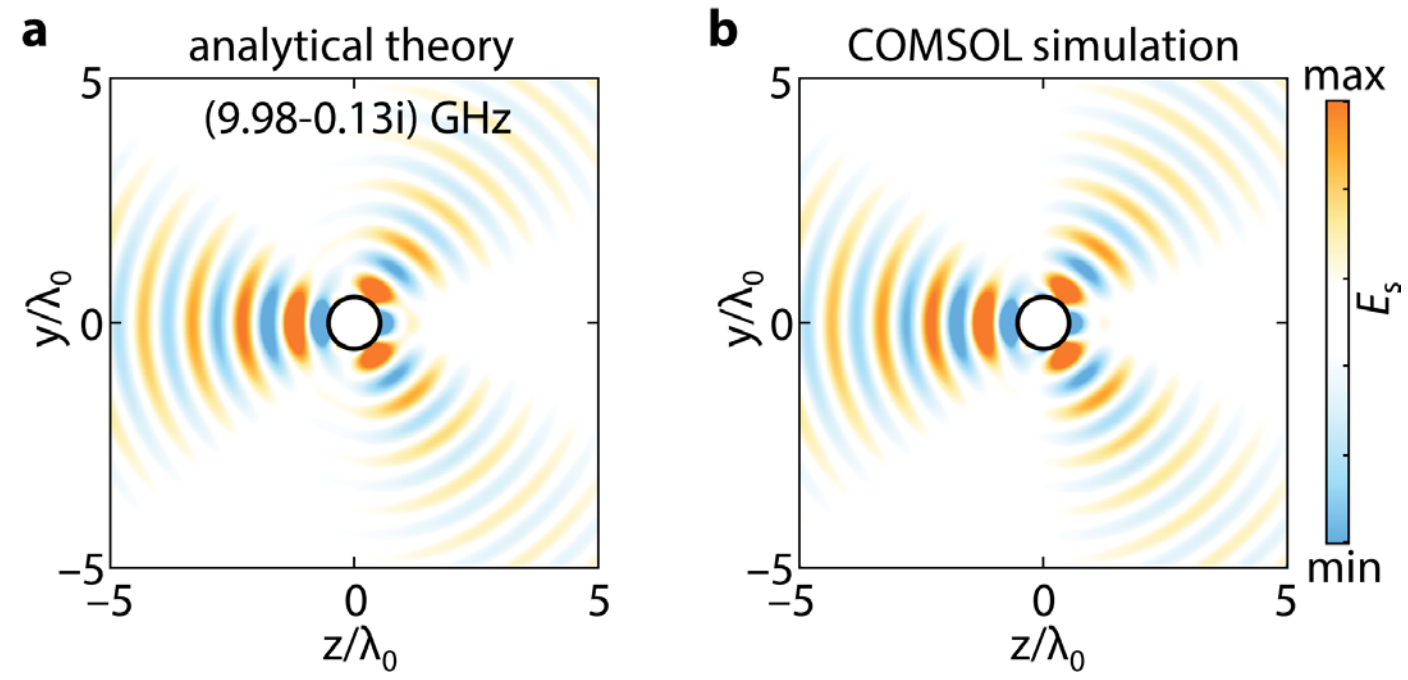

**Fig. S1 | Complex-frequency scattered field distribution. a**, Analytical simulation. **b**, Frequency-domain COMSOL simulation. The structure setup is identical to that of the main text.

### S1.1.2 Radiation pattern

Here, we derive the radiation pattern and normalized forward scattering cross section extracted directly from the measured field distribution. This is highly practical in our experimental setup, which directly provides access to the field distribution. Let us start with the wavevector-spectrum representation of the scattered field[S3], namely

$$E_{\mathrm{s}}(x,y,z)=\int_{-\infty}^{\infty}\mathrm{d}k_y\int_{-\infty}^{\infty}\mathrm{d}k_z\;\tilde{E}_{\mathrm{s}}\left(k_y,k_z\right)e^{ik_yy+ik_zz+ik_xx}\tag{S6}$$

For the far-field approximation of interest, the far-field scattered field $E_{\mathrm{s},\infty}$ is given by

$$E_{\mathrm{s},\infty}(x,y,z)=\iint_{k_y^2+k_z^2\leq k^2}\mathrm{d}k_y\mathrm{d}k_z\;\tilde{E}_{\mathrm{s}}\left(k_y,k_z\right)e^{ik_yy+ik_zz+ik_xx}\tag{S7}$$

where the integration range of $k_y^2+k_z^2\leq k^2$ is used to exclude the evanescent waves. Equation (S7) can be evaluated by stationary phase method as $r\rightarrow\infty$[S3], namely

$$E_{\mathrm{s},\infty}(x,y,z)\propto\tilde{E}_{\mathrm{s}}\left(k_y,k_z\right)\frac{e^{ikr}}{r}\tag{S8}$$

Physically, equation (S8) means that $E_{\mathrm{s},\infty}(x,y,z)$ is contributed by only the plane wave that propagates towards it, namely that the wavevector $\overline{k}=(k_x,k_y,k_z)$ is parallel to the observation direction $\overline{r}=(x,y,z)$. Furthermore, using the relation between $\sigma_{\mathrm{sca}}^{\mathrm{d}}(\theta)$ and $E_{\mathrm{s},\infty}(x=0,y,z)$[S1,S2], one has

$$\sigma_{\mathrm{sca}}^{\mathrm{d}}(\theta)=\lim_{r\rightarrow\infty}4\pi r^2\left|E_{\mathrm{s},\infty}(x=0,y,z)\right|^2/|E_0|^2\propto\left|\tilde{E}_{\mathrm{s}}(k\cos\theta\,,k\sin\theta)\right|^2/|E_0|^2\tag{S9}$$

In practice, $\tilde{E}_{\mathrm{s}} = \frac{1}{(2\pi)^2} \iint_A \mathrm{d}y\, \mathrm{d}z\, E_{\mathrm{s}}(0, y, z) e^{-ik_y y - ik_z z}$ in equation (S9) is obtained in a based on a integration region of $A \coloneqq \{(y,z) | y, z \in [-5\lambda_0, 5\lambda_0]\} \backslash \{(y,z) | y^2 + z^2 \leq \lambda_0^2\}$.

As shown in Fig. S2, equation (S9) can be verified by an analytical benchmark[S1,S2], namely

$$\sigma_{\mathrm{sca}}^{\mathrm{d}}(\theta) = \frac{4\pi}{|k|^2} \sum_{n=1}^{\infty} \frac{(2n+1)}{n(n+1)} \left( a_n(\omega) \pi_n(\cos\theta) + b_n(\omega) \tau_n(\cos\theta) \right) \tag{S10}$$

where $a_n(\omega)$ and $b_n(\omega)$ are given in equations (S4) and (S5).

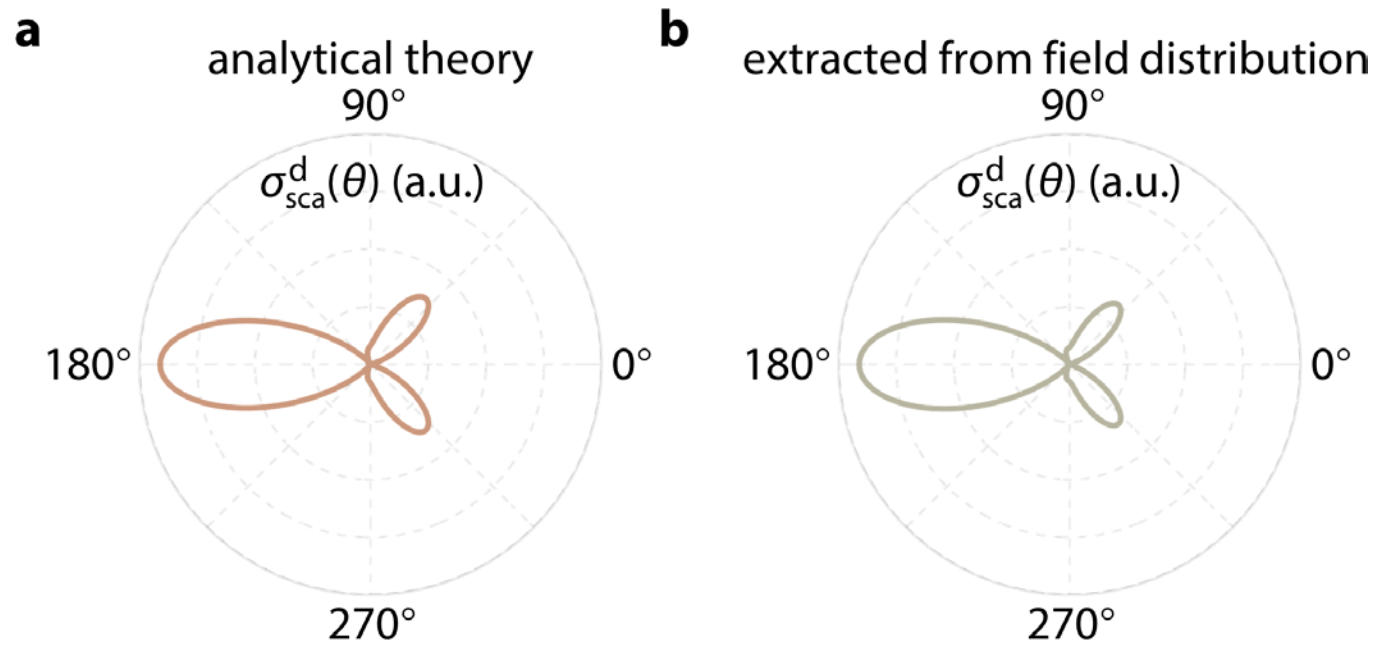

**Fig. S2 | Complex-frequency radiation pattern. a**, Analytical theory. **b**, Extracted from field distribution. For illustration here and below, $\sigma_{\mathrm{sca}}^{\mathrm{d}}(\theta)$ is normalized to $\max\left[\sigma_{\mathrm{sca}}^{\mathrm{d}}(\theta), \theta\right]$ for each pattern.

## S1.2 Forward scattering response in the complex frequency plane

In this subsection, we map the forward scattering responses in the complex-frequency plane, to identify a series of forward-scattering zeros and poles in the lower complex-frequency plane (Fig. S3).

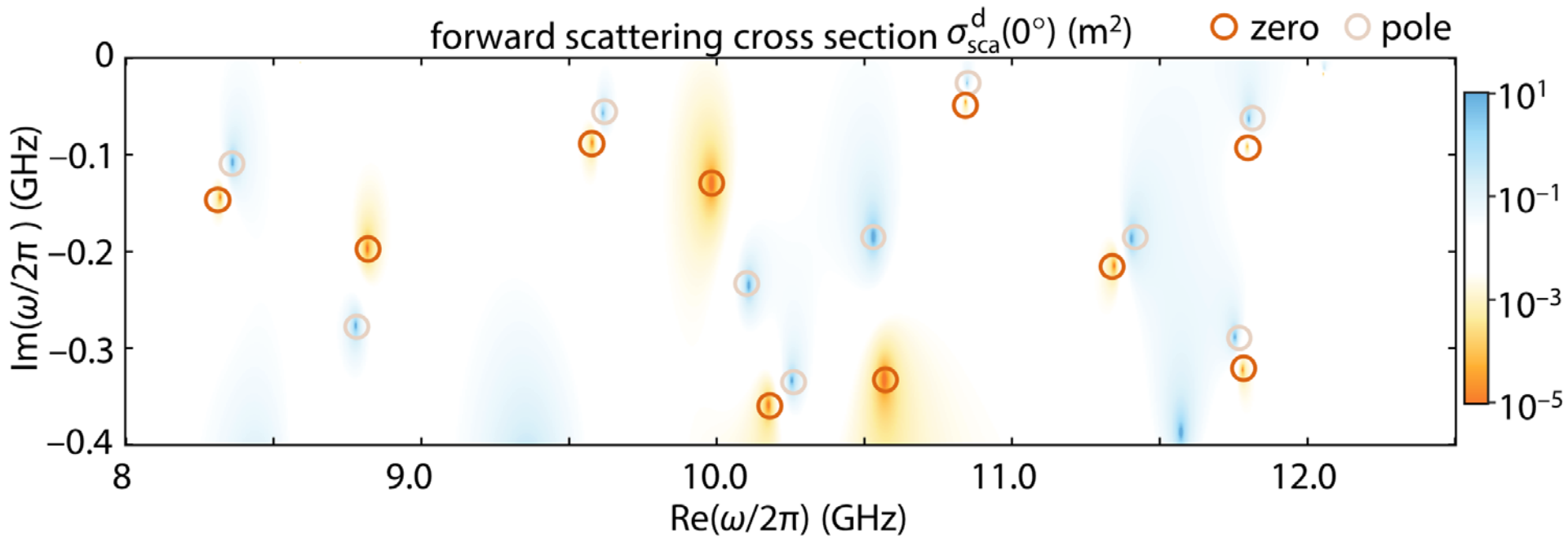

**Fig. S3 | Forward scattering zeros and poles in the complex-frequency plane.** The results here are analytically obtained via equation (S10).

## S2 More discussion on the artificial pole for zero-forward Kerker scattering

In this section, we provide more discussion on the artificial pole for zero-forward Kerker scattering.

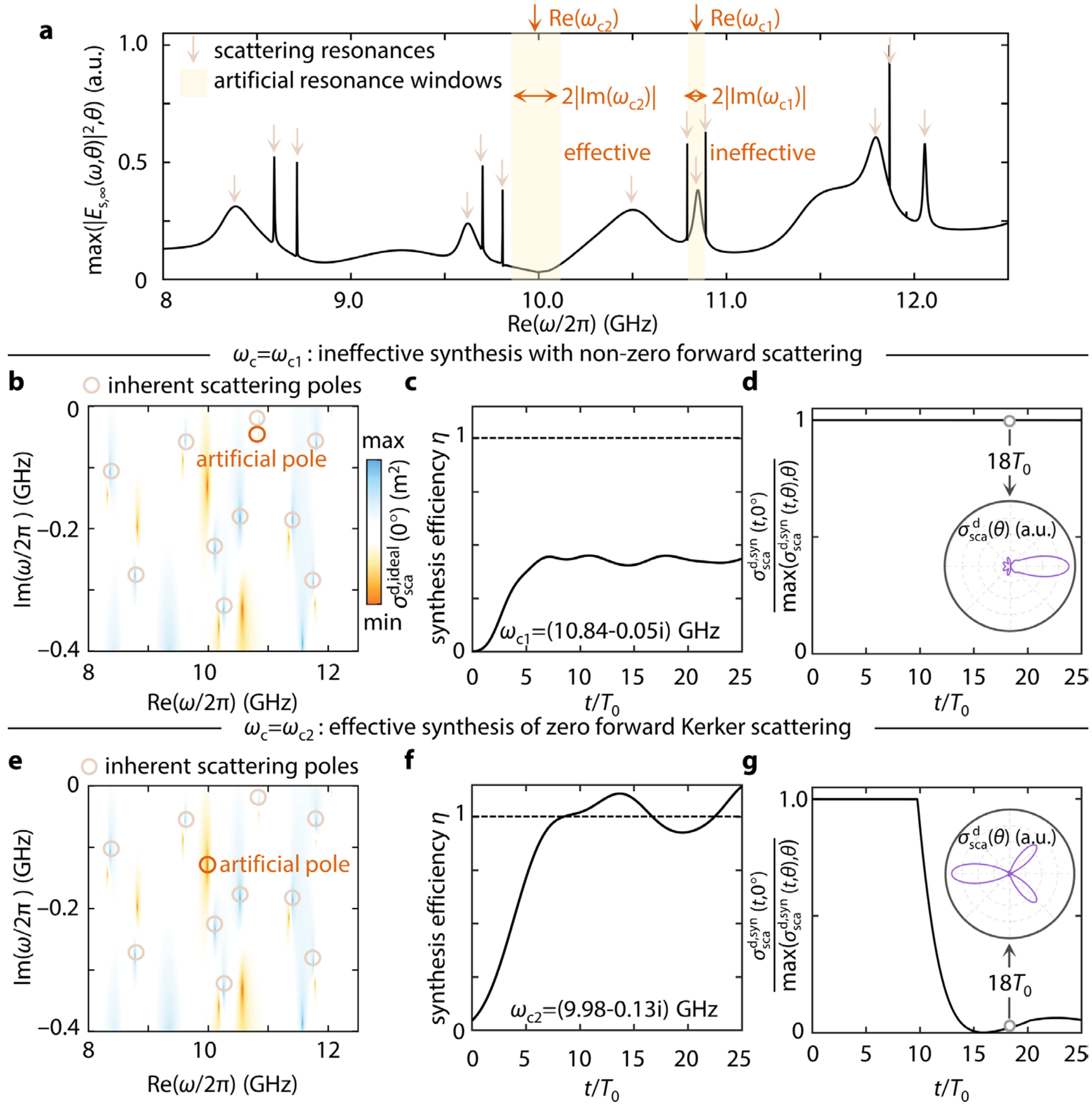

**Fig. S4 | Artificial pole for zero-forward Kerker scattering. a**, Maximal far-field scattering intensity spectrum $\max\left(\left|E_{s,\infty}(\omega,\theta)\right|^2,\theta\right)$. **b,e**, Different artificial poles created at the position of forward-scattering zeros. **c,f**, Synthesis efficiency for different artificial poles $\eta=\int_0^{2\pi}\mathrm{d}\theta\ \sigma_{\mathrm{sca}}^{\mathrm{d,syn}}(\theta)\Big/\int_0^{2\pi}\mathrm{d}\theta\ \sigma_{\mathrm{sca}}^{\mathrm{d,ideal}}(\theta)$, where $\sigma_{\mathrm{sca}}^{\mathrm{d,syn}}(\theta)$ and $\sigma_{\mathrm{sca}}^{\mathrm{d,ideal}}(\theta)$ are the differential scattering cross sections of the synthesized and ideal complex-frequency fields, respectively. **d,g**, Evolution of the forward scattering cross section of the synthesized field over time. For illustration, the far-filed observation point in **a** is $r=10\lambda_0$ and $\phi=90°$. Before introducing

the artificial pole, the scattering spectrum in **a** is fully captured by the distribution of inherent scattering zeros and poles in the complex-frequency plane.

Figure S4 provides a useful rule for creating an artificial pole, by simply studying the frequency spectrum of the scattering system before the artificial pole is introduced. The candidate artificial poles $\omega_{\mathrm{c}} = \mathrm{Re}(\omega_{\mathrm{c}}) - i \cdot |\mathrm{Im}(\omega_{\mathrm{c}})|$ manifest themselves as resonance windows characterized by a central frequency $\mathrm{Re}(\omega_{\mathrm{c}})$ and a bandwidth $2|\mathrm{Im}(\omega_{\mathrm{c}})|$ (Fig. S4a). For the artificial pole whose artificial resonance (e.g., $\omega_{\mathrm{c1}}$ in Fig. S4a,b) spectrally overlaps with inherent scattering resonances, the synthesis is ineffective with non-zero forward scattering (Fig. S4c,d). In contrast, by creating an artificial pole whose resonance (e.g., $\omega_{\mathrm{c2}}$ in Fig. S4a,e) does not spectrally overlap with any pronounced scattering resonances, the intrinsic scattering poles are suppressed by the factor of $1/(\omega - \omega_{\mathrm{c}})$ and the zero-forward Kerker scattering is successfully synthesized (Fig. S4f,g).

## S3 Synthesized zero-forward Kerker scattering by a cylinder

In this section, we apply the synthetic framework to a two-dimensional cylindrical scattering system. Consider a cylindrical scatterer with a radius $a$, a length $l \gg \lambda_0$, and aligned with the $z$ axis of $(\rho, \theta, z)$ cylindrical coordinates (Fig. S5a). The incident plane wave travelling along $x$ direction has an electric field of $\overline{E}_{\mathrm{inc}} = \hat{z} \cdot E_0 e^{ikx - i\omega t}$, where $k = \omega/c$. The corresponding scattered field is given by[1]:

$$\overline{E}_{\mathrm{s}} = \hat{z} \cdot E_{\mathrm{s}} e^{-i\omega t} = \hat{z} \cdot E_0 \sum_{n=-\infty}^{\infty} i^n a_n H_n^{(1)}(k\rho) e^{in\theta}\, e^{-i\omega t} \tag{S11}$$

$$a_n(\omega) = i^n \frac{\sqrt{\varepsilon_{\mathrm{r}}/\mu_{\mathrm{r}}} J_n(ka) J_n'(k_{\mathrm{s}}a) - J_n'(ka) J_n(k_{\mathrm{s}}a)}{\sqrt{\varepsilon_{\mathrm{r}}/\mu_{\mathrm{r}}} H_n^{(1)}(ka) J_n'(k_{\mathrm{s}}a) - H_n^{(1)\prime}(ka) J_n(k_{\mathrm{s}}a)} \tag{S12}$$

where $H_n^{(1)}$ and $J_n$ denotes the Hankel function and the Bessel function of the first kind, respectively; $\varepsilon_{\mathrm{r}}$ and $\mu_{\mathrm{r}}$ are the relative permittivity and permeability of the cylinder. The differential scattering cross section is defined and evaluated as

$$\sigma_{\mathrm{sca}}^{\mathrm{d}}(\theta) = \mathrm{d}\sigma/\mathrm{d}\theta \propto \left|\tilde{E}_{\mathrm{s}}(k\cos\theta, k\sin\theta)\right|^2 / |E_0|^2 \tag{S13}$$

In practice, $\tilde{E}_{\mathrm{s}}(k_x, k_y) = \frac{1}{(2\pi)^2} \iint_A \mathrm{d}x\, \mathrm{d}y\, E_{\mathrm{s}}(x, y, 0) e^{-ik_x x - ik_y y}$, where $A \coloneqq \{(x, y) | x, y \in [-5\lambda_0, 5\lambda_0]\} \setminus \left\{(x, y) | x^2 + y^2 \le \lambda_0^2\right\}$.

Since the real-frequency measurements of cylindrical scattering theory are accurate (Fig. S5b), zero-forward Kerker scattering by a cylinder at a synthesized complex frequency can be effectively synthesized, as shown in Fig. S5c.

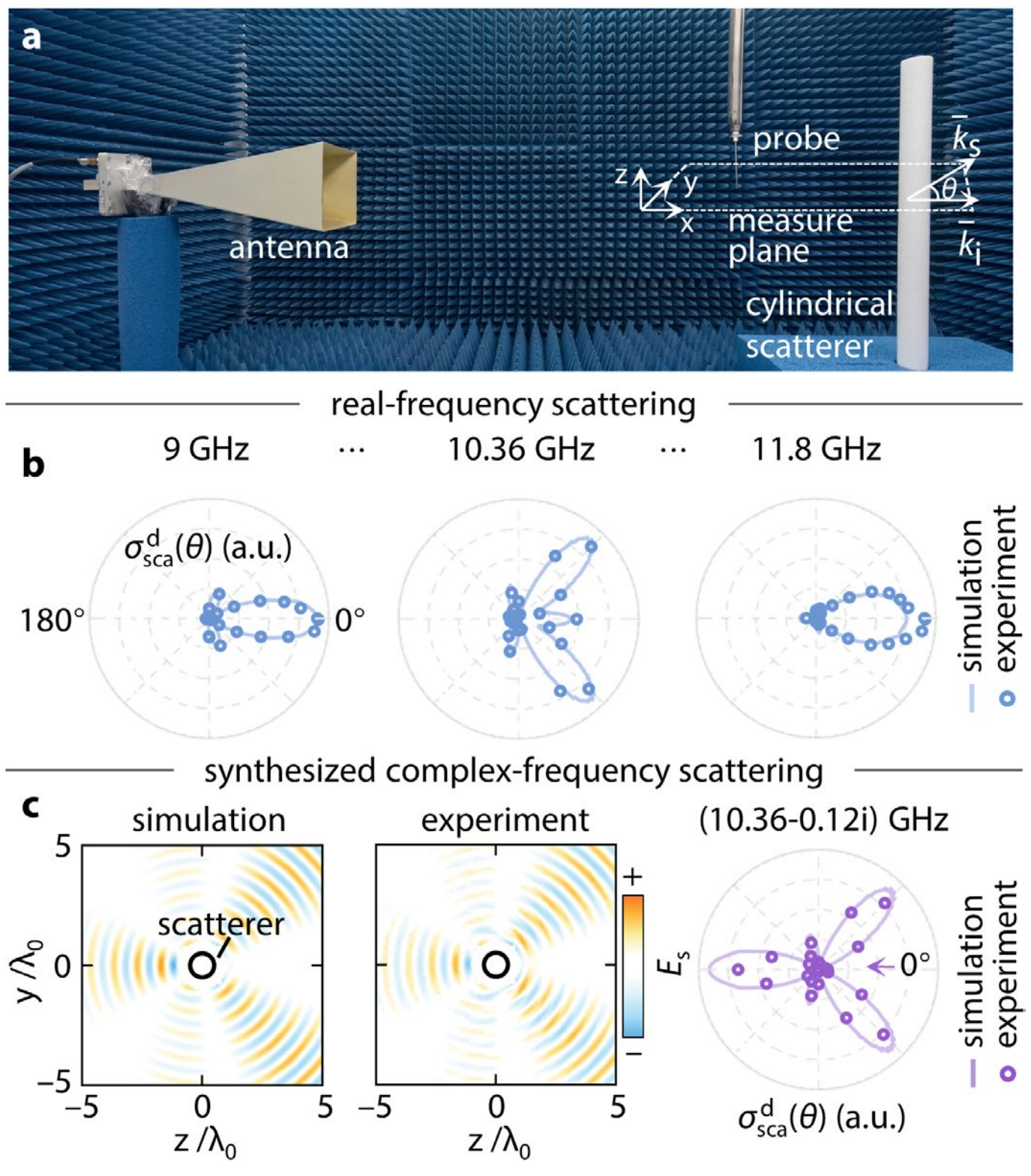

**Fig. S5 | Experimental realization of the zero-forward Kerker scattering by a cylinder. a**, Experimental setup. **b**,**c**, Simulated and measured results at real frequencies and a synthesized complex frequency. For illustration, the scatterer has a radius of $a = 15$ mm and a relative permittivity of $\varepsilon_\mathrm{r} = 9.5$; the real-frequency measurement is conducted in the frequency range 9.0~11.8 GHz and with a frequency step of 0.05 GHz.

# S4 More discussion on synthesized zero-forward Kerker scattering

## S4.1 Relationship among scattering responses under real-frequency and different types of complex-frequency excitation

In this subsection, we clarify the relationship among real-frequency scattering responses and the different complex-frequency responses illustrated in Fig. S6. Specifically, we note that the synthesized scattering responses (e.g., Fig. S6b) are numerically reconstructed from real-frequency measurements (e.g., Fig. S6a). Importantly, we emphasize that the numerically synthesized scattering response in Fig. S6b is precisely equivalent to that obtained under physical complex-frequency excitation in Fig. S6c, where a temporal truncation of an ideal complex-frequency incident wave is necessary to avoid an unphysical divergence as $t \to -\infty$. Moreover, both the synthesized and physical scattering responses in Fig. S6b,c are in good agreement with the response under ideal complex-frequency excitation shown in Fig. S6d.

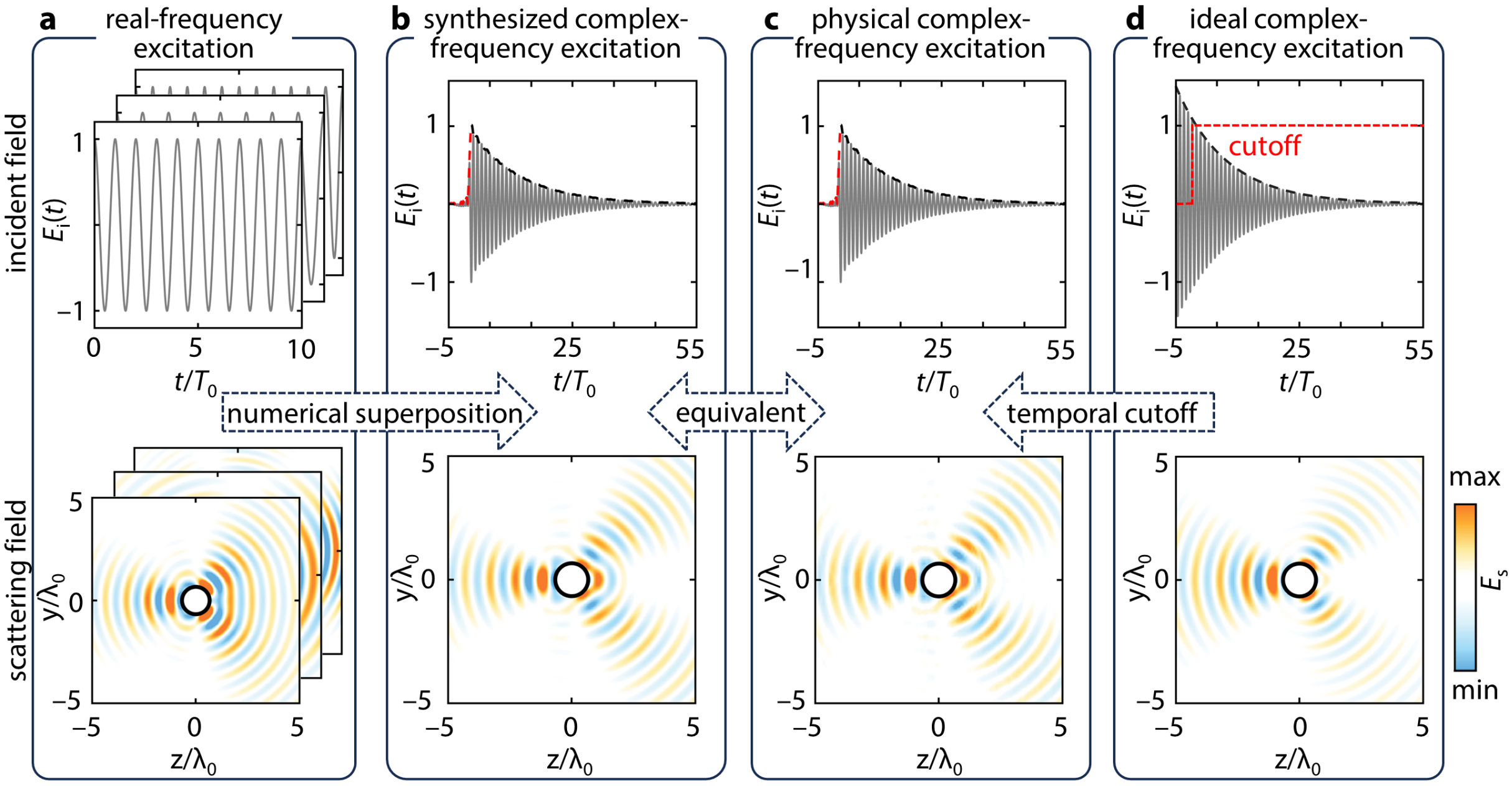

**Fig. S6 | Relationship among scattering responses under real-frequency and different types of complex-frequency excitation.** We compare the scattering responses under real-frequency excitation in **a**, synthesized complex-frequency excitation in **b**, physical complex-frequency excitation in **c**, and ideal complex-frequency excitation in **d**. The upper panels show the incident-wave profiles, while the lower panels show the corresponding scattering responses. Importantly, synthesized complex-frequency response in **b** is equivalent to the physical complex-frequency response in **c**. The structural setup is the same as that in Fig. 2 of the main text, except that a frequency range of 2~18 GHz and a corresponding synthesis bandwidth of 16 GHz are used to make the temporal cutoff visually clearer.

## S4.2 Multipolar decomposition of synthesized zero-forward Kerker scattering

In this subsection, we further analyze the synthesized zero-forward scattering condition through a multipole expansion[S4]. As noted by the reviewer, synthesized zero-forward Kerker scattering arises from the destructive interference among electric and magentic multipoles. In contrast to the conventional zero-forward Kerker condition, where electric and magnetic dipoles ($n = 1$) are dominant and out of phase, we highlight in Fig. S7a that the synthesized zero-forward Kerker condition relies on higher-order multipoles (e.g., those with order $n = 1, 2, 3, 4$). To be specific, destructive interference of these multipoles leads to vanishing scattering amplitude (Fig. S7b-c) and differential cross section (Fig. S7d) in the forward direction.

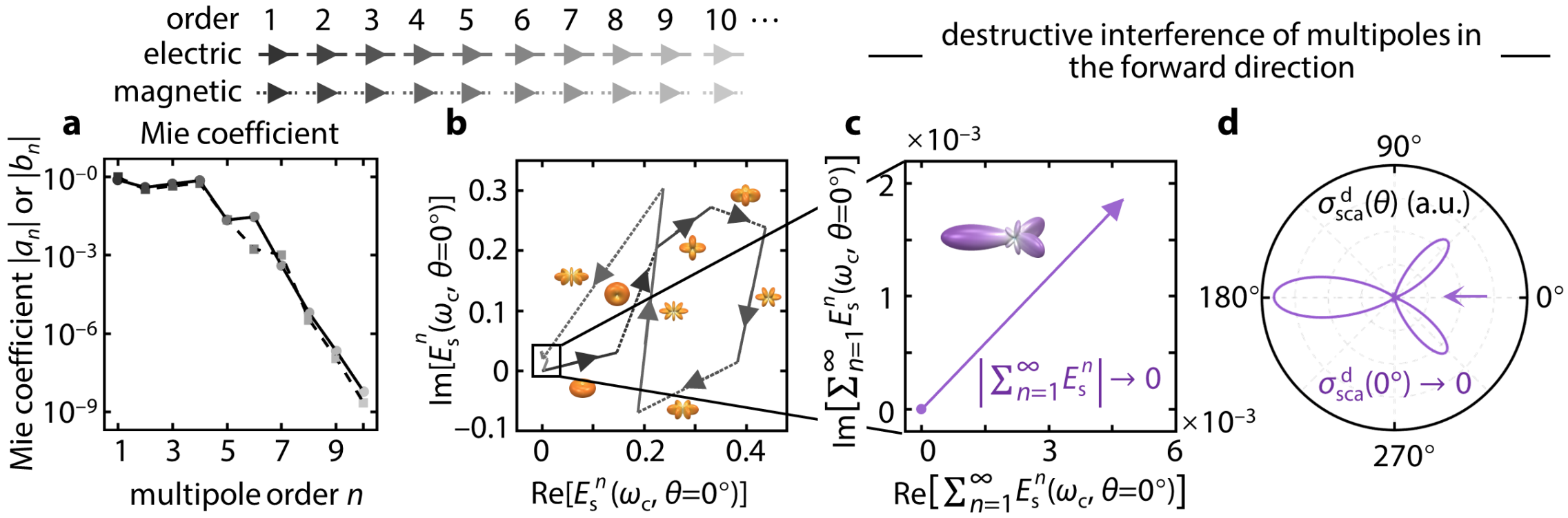

**Fig. S7 | Multipolar decomposition of synthesized zero-forward Kerker scattering.** Absolute scattering coefficients $|a_n|$ and $|b_n|$ of electric and magnetic multipoles. Multipolar cofficients $a_n$ and $b_n$ are obtained by projecting the scattered field onto the corresponding vector spherical harmonics[S5]. **b,** Multipolar forward scattering amplitudes $E_{\mathrm{s}}^{n}(\omega_{\mathrm{c}}, \theta = 0°)$. **c**, Total forward scattering amplitude $\sum_{n=1}^{\infty} E_{\mathrm{s}}^{n}(\omega_{\mathrm{c}}, \theta = 0°)$. All scattering amplitudes in **b,c** are mapped as vectors in the complex plane of eletric field. **d,** Normalized differential scattering cross section $\sigma_{\mathrm{sca}}^{\mathrm{d}}(\theta)$ for synthesized zero-forward scattering. The structural setup is the same as Fig. 2 of the main text.

### S4.3 Sensitivity to synthesis frequency, synthesis bandwidth and frequency step

In this subsection, we provide a quantitative estimate of this error and discuss sensitivity to the chosen $\omega_{\mathrm{c}}$, finite measurement bandwidth, and frequency step. In fact, the synthesis error, quantified by the ratio of the residual contribution from the inherent scattering poles to that from the artificial pole, can be reduced to a rather low level, as shown in Fig. S8. Specifically, the synthesis error can be suppressed to $\Delta \leq 10\%$ by choosing a proper synthesis frequency $\omega_{\mathrm{c}}$, together with a sufficiently broad synthesis bandwidth, e.g., $\omega_{\mathrm{B}}/\mathrm{Re}(\omega_{\mathrm{c}}) \gtrsim 0.15$, and a small frequency step, e.g., $\Delta\omega/\mathrm{Re}(\omega_{\mathrm{c}}) \lesssim 0.02$.

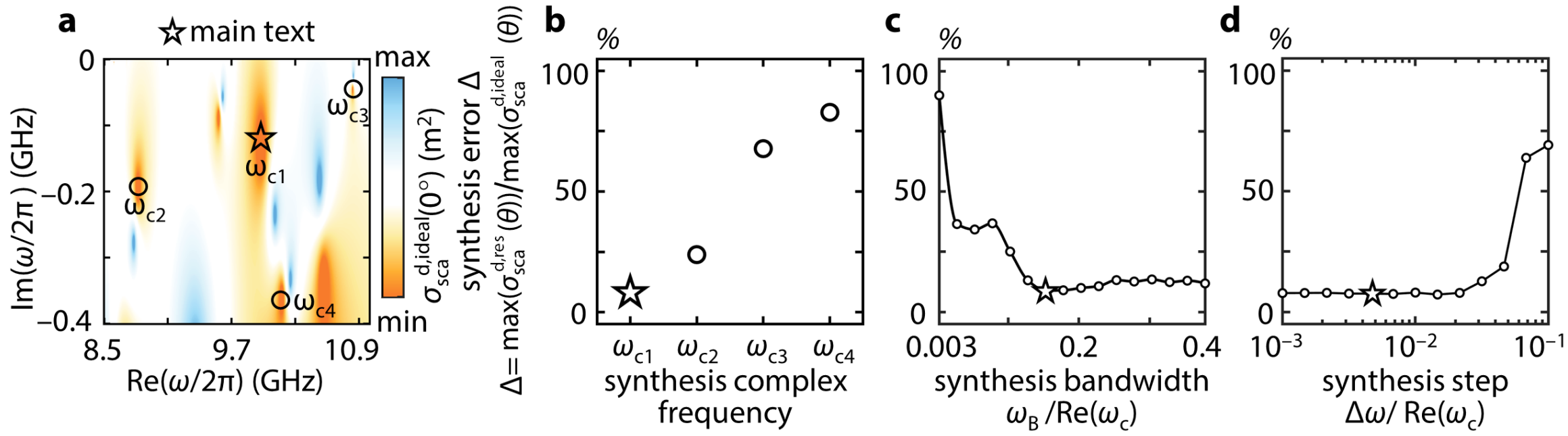

**Fig. S8 | Sensitivity of zero-forward Kerker scattering to synthesis frequency, synthesis bandwidth and frequency step. a,** Forward scattering cross section mapped in the complex-frequency plane. **b-d**,

Dependence of synthesis error on synthesis frequency, synthesis bandwidth and frequency step. The synthesis error is defined as the ratio of the residual contribution from the intrinsic scattering poles to that from the artificial pole, namely, $\Delta = \max\left[\sigma_{\mathrm{sca}}^{\mathrm{d,res}}(\theta), \theta\right] / \max\left[\sigma_{\mathrm{sca}}^{\mathrm{d,ideal}}(\theta), \theta\right]$. The structural setup is the same as that in Fig. 2 of the main text.

## S4.4 Experimental error estimates

In this subsection, we carefully examine the potential error sources for the experimental conditions adopted in this work. Specifically, in Fig. S9, we quantify the minimum experimental requirements guaranteeing the convergence of error, namely a probe-scanning resolution of $\delta r_{\mathrm{p}} \lesssim 0.33\lambda_0$, a minimum scan area of $L \gtrsim 3.33\lambda_0$, and an antenna directivity $D \gtrsim 10.6$ dBi. As a result, a vanishing scattering cross section is synthesized, e.g., $\sigma_{\mathrm{sca}}^{\mathrm{d}}(0°)/\max\left[\sigma_{\mathrm{sca}}^{\mathrm{d}}(\theta), \theta\right] \lesssim 1.2\%$. This also indicates that additional experimental uncertainties, such as background reflections, are readily suppressed in our experimental setup.

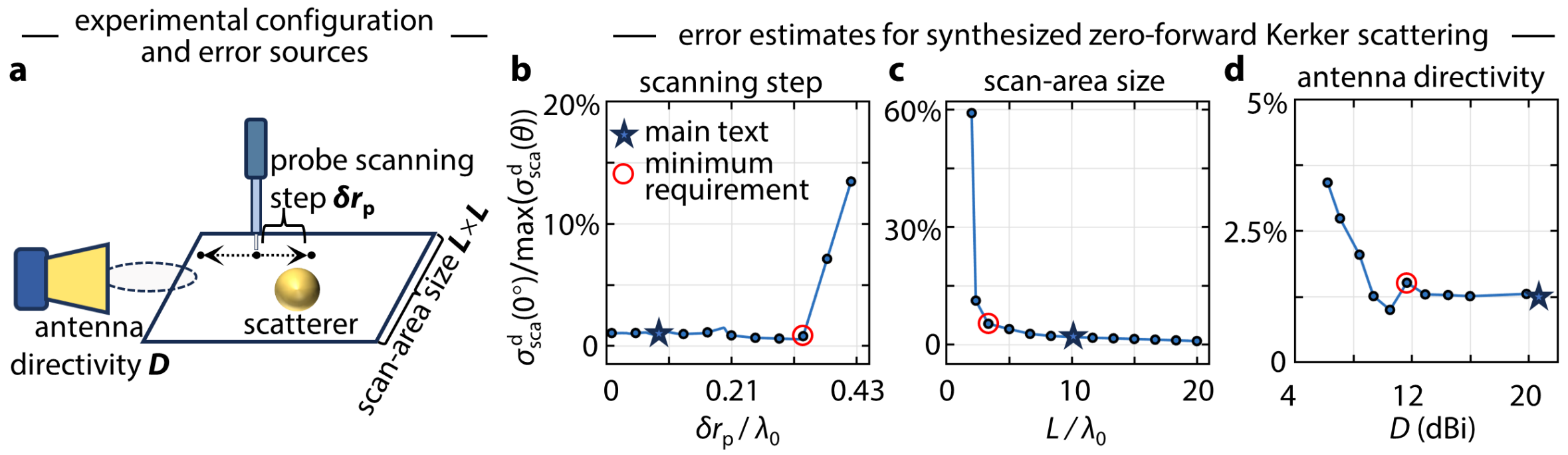

**Fig. S9 | Experimental error estimates for zero-forward Kerker scattering via complex-frequency excitation. a,** Schematic of the experimental configuration and error sources. Potential error sources include probe-scanning resolution $\delta r_{\mathrm{p}}$, finite scan area $L$, and antenna directivity $D$. **b-d,** Normalized forward scattering cross section $\sigma_{\mathrm{sca}}^{\mathrm{d}}(0°)/\max\left[\sigma_{\mathrm{sca}}^{\mathrm{d}}(\theta), \theta\right]$ as a function of various experimental error sources. In each panel, only the indicated parameter is varied, with all other parameters held fixed at their main-text values. The other structural setup is the same as Fig. 3 of the main text, and the antenna is positioned at a distance of 420 mm from the scatterer.

## S4.5 Influence of synthesis frequency, scatterer properties, and material loss

In this subsection, we show in Figs. S10-S12 the influence of synthesis frequency, scatterer properties, and material loss on synthesized zero-forward Kerker scattering.

In fact, the synthesized zero-forward Kerker scattering is robust against variations in the scatterer properties, provided that the synthesis frequency is appropriately chosen for each scatterer. First, we highlight that the synthesis error can be suppressed to a relatively low level by properly selecting the synthesis frequency (Fig. S10a,b), yet not every forward-scattering zero corresponds to a valid synthesis frequency (Fig. S10c-g; see also supplementary section S2 and Fig. S4). More importantly, indeed, we verify in Fig. S11 that a low synthesis error can be maintained for randomly chosen scatterers across the diameter-permittivity parameter space.

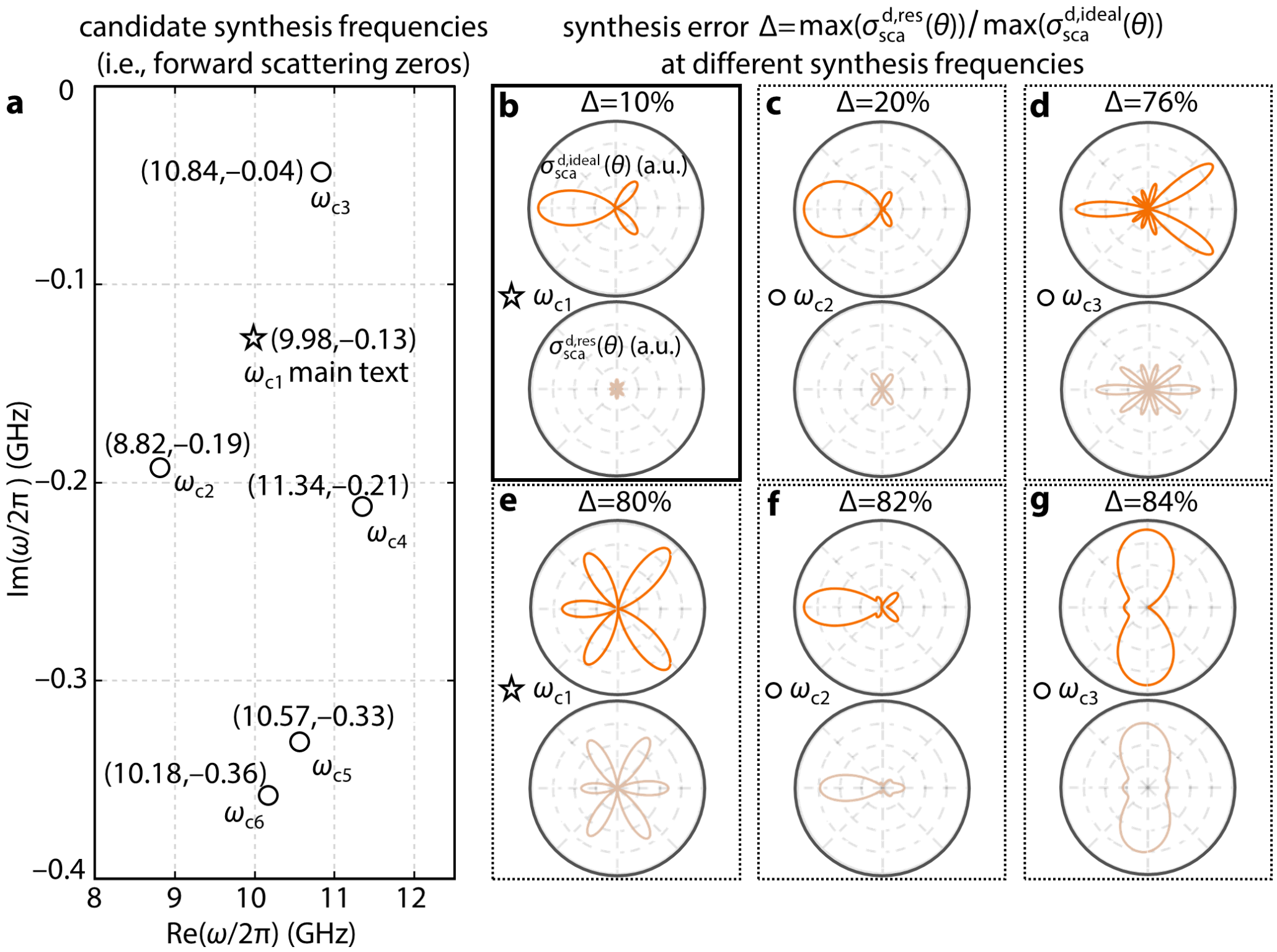

**Fig. S10 | Influence of synthesis frequency on zero-forward Kerker scattering. a,** Candidate synthesis frequencies $\omega_{ci}$ ($i = 1,2\dots 6$) mapped in the complex-frequency plane. **b-g**, Differential scattering cross sections for various synthesis frequencies. In each panel, the upper subpanel shows the ideal complex-frequency scattering response $\sigma_{\text{sca}}^{\text{d,ideal}}(\theta)$, while the lower subpanel shows the residual contribution from the inherent scattering poles $\sigma_{\text{sca}}^{\text{d,res}}(\theta)$. The synthesis error is effectively suppressed by carefully choosing the synthesis frequency; for example, for $\omega_{\text{c1}}$ used in the main text, $\Delta = \max\left[\sigma_{\text{sca}}^{\text{d,res}}(\theta), \theta\right] / \max\left[\sigma_{\text{sca}}^{\text{d,ideal}}(\theta), \theta\right] \lesssim 10\%$.

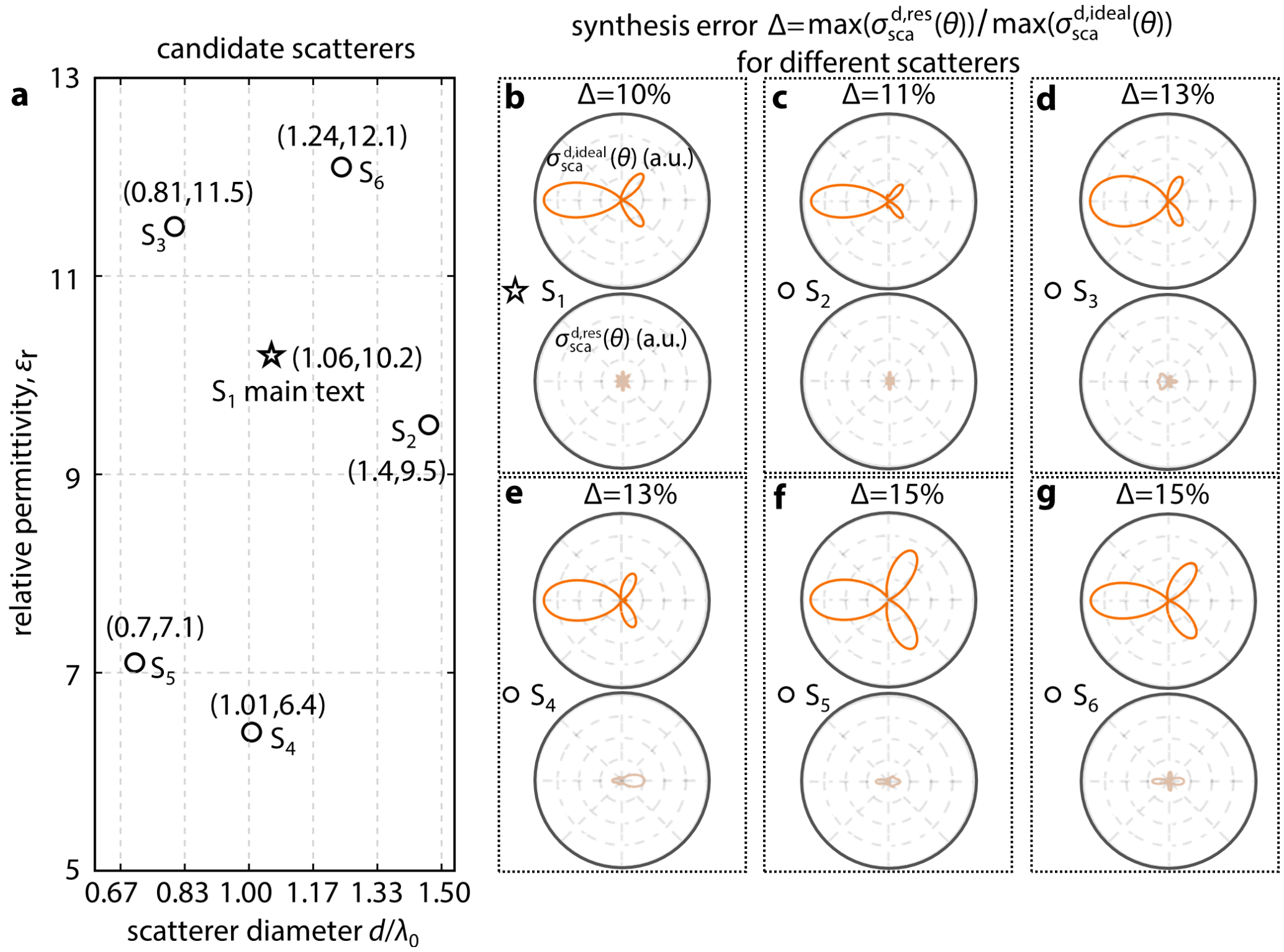

**Fig. S11 | Robustness of synthesized zero-forward Kerker scattering against scatterer properties. a,** Candidate spherical scatterers $S_i$ in the parameter space of scatterer diameter and relative permittivity. **b-g**, Normalized differential scattering cross sections for candidate scatterers. In each panel, the upper subpanel shows the ideal complex-frequency scattering response $\sigma_{sca}^{d,ideal}(\theta)$, while the lower subpanel shows the residual contribution from the inherent scattering poles $\sigma_{sca}^{d,res}(\theta)$. Importantly, all six randomly chosen scatterers with parameters denoted in **a** exhibit vanishing forward scattering amplitudes and small synthesis errors in **b-g**, namely $\Delta \lesssim 15\%$, for $\forall S_i$ $(i = 1,2\ldots6)$. The synthesis complex frequencies for the six scatterers are $(9.98 - i0.13)$ GHz, $(7.48 - i0.1)$ GHz, $(10.88 - i0.18)$ GHz, $(7.53 - i0.14)$ GHz, $(10.21 - i0.15)$ GHz and $(7.79 - i0.13)$ GHz, respectively.

Furthermore, we note that our findings exhibit a certain degree of tolerance to material loss. As the material loss increases, the inherent scattering poles and zeros shift along the imaginary-frequency axis, as shown in Fig. S12a,c,e. Notably, for reasonably small material losses, e.g., a loss tangent of $\tan\delta \leq 5\%$, the corresponding zero-forward Kerker scattering can still be effectively recovered by simply employing a larger synthesis gain, i.e., a synthesis complex frequency with a larger imaginary part.

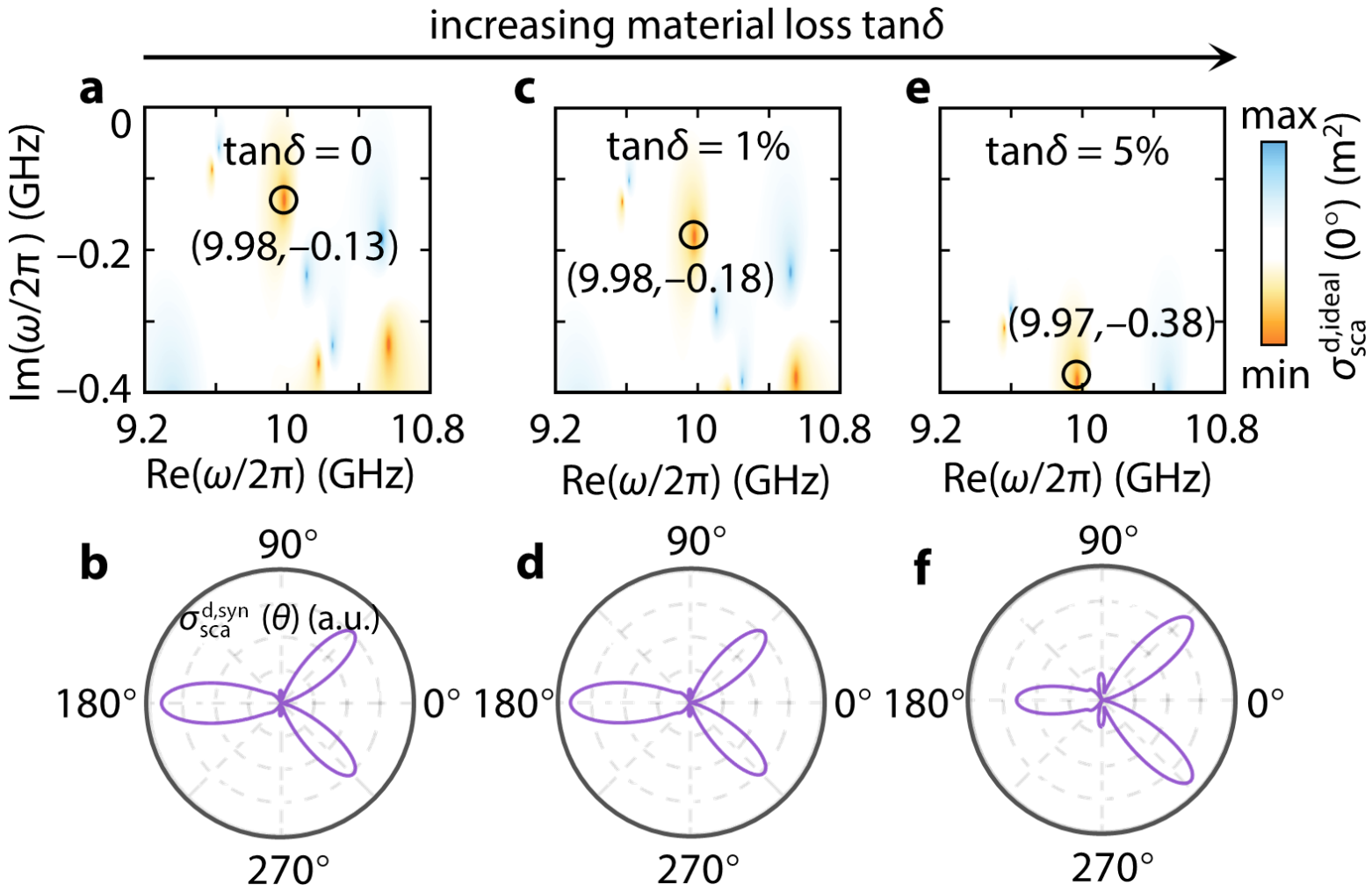

**FIG. S12 | Influence of material loss on zero-forward Kerker scattering via synthetic complex-frequency excitation. a,c,e,** Forward scattering cross section mapped in the complex-frequency plane. **b**,**d**,**f**, Normalized differential cross sections for the synthesized complex-frequency responses. The structural setup is the same as Fig. 2 of the main text, except that the loss tangent are 0 in **a,b**, 1% in **c,d**, and 5% in **e,f**, respectively.